\documentclass[a4paper, 12pt]{article}

\usepackage{graphicx}
\usepackage{amsmath}
\usepackage{amssymb}
\usepackage{authblk}
\usepackage[a4paper]{geometry}

\title{Description of CRESST-II data}

\date{\today}

\author[1]{\small G.~Angloher}
\author[1]{P.~Bauer}
\author[1,a]{A.~Bento}
\author[2]{C.~Bucci}
\author[2,b]{L.~Canonica}
\author[3]{X.~Defay}
\author[3,c]{A.~Erb}
\author[3]{F.~v.~Feilitzsch}
\author[1]{N.~Ferreiro~Iachellini}
\author[2]{P.~Gorla}
\author[4,5]{A.~G\"utlein\thanks{Corresponding author: achim.guetlein@oeaw.ac.at}}
\author[1]{D.~Hauff}
\author[6]{J.~Jochum}
\author[1]{M.~Kiefer}
\author[1]{C.~Kistner}
\author[4,5]{H.~Kluck}
\author[7]{H.~Kraus}
\author[3]{J.-C.~Lanfranchi}
\author[6]{J.~Loebell}
\author[1]{M.~Mancuso}
\author[3]{A.~M\"unster}
\author[2]{C.~Pagliarone}
\author[1]{F.~Petricca}
\author[3]{W.~Potzel}
\author[1]{F.~Pr\"obst}
\author[4,5]{R.~Puig}
\author[1]{F.~Reindl\thanks{Current address: INFN - Sezione di Roma, I-00185 Roma, Italy}}
\author[3]{S.~Roth}
\author[6]{K.~Rottler}
\author[6]{C.~Sailer}
\author[2,d]{K.~Sch\"affner}
\author[4,5]{J.~Schieck}
\author[1]{J.~Schmaler}
\author[6]{S.~Scholl}
\author[3]{S.~Sch\"onert}
\author[1]{W.~Seidel}
\author[3]{M.v.~Sivers}
\author[1]{L.~Stodolsky}
\author[6]{C.~Strandhagen}
\author[1]{R.~Strauss}
\author[1]{A.~Tanzke}
\author[3]{H.H.~Trinh~Thi}
\author[4,5]{C.~T\"urko\v{g}lu}
\author[6]{M.~Uffinger}
\author[3]{A.~Ulrich}
\author[6]{I.~Usherov}
\author[3]{S.~Wawoczny}
\author[3]{M.~Willers}
\author[1]{M.~W\"ustrich}
\author[3]{A.~Z\"oller}

\affil[1]{\footnotesize Max-Planck-Institut f\"ur Physik, D-80805 M\"unchen, Germany}
\affil[2]{INFN, Laboratori Nazionali del Gran Sasso, I-67010 Assergi, Italy}
\affil[3]{Physik-Department and Excellence Cluster Universe, Technische Universit\"at M\"unchen, D-85747 Garching, Germany}
\affil[4]{Institut f\"ur Hochenergiephysik der \"Osterreichischen Akademie der Wissenschaften, A-1050 Wien, Austria}
\affil[5]{Atominstitut, Vienna University of Technology, A-1020 Wien, Austria}
\affil[6]{Eberhard-Karls-Universit\"at T\"ubingen, D-72076 T\"ubingen, Germany}
\affil[7]{Department of Physics, University of Oxford, Oxford OX1 3RH, United Kingdom}
\affil[a]{Also at: Departamento de Fisica, Universidade de Coimbra, P3004 516 Coimbra, Portugal}
\affil[b]{Also at: Massachusetts Institute of Technology, Cambridge, MA 02139, USA}
\affil[c]{Also at: Walther-Mei\ss{}ner-Institut f\"ur Tieftemperaturforschung, D-85748 Garching, Germany}
\affil[d]{Also at: GSSI-Gran Sasso Science Institute, 67100, L'Aquila, Italy}

\begin{document}

\maketitle

\begin{abstract}
In Phase 2 of CRESST-II 18 detector modules were operated for about two years (July 2013 - August 2015). Together with this document we are publishing data from two detector modules which have been used for direct dark-matter searches. With these data-sets we were able to set world-leading limits on the cross section for spin-independent elastic scattering of dark matter particles off nuclei. We publish the energies of all events within the acceptance regions for dark-matter searches. In addition, we also publish the energies of the events within the electron-recoil band. This data set can be used to study interactions with electrons of CaWO$_4$. In this document we describe how to use these data sets. In particular, we explain the cut-survival probabilities required for comparisons of models with the data sets.  
\end{abstract}

\section{Data selection}

CRESST-II detector modules consist of two cryogenic detectors: The phonon detector based on a CaWO$_4$ crystal measures the phonons generated by the energy deposition from an interaction of a particle with the CaWO$_4$ crystal. The light detector based on a silicon-on-sapphire disc measures the scintillation light generated  simultaneously in the CaWO$_4$ crystal, see e.g. \cite{CresstDetectors_1} for further details.

The published data described in this document were obtained by two detector modules: TUM40 \cite{CresstDetectors_1},\cite{TUM40} and Lise \cite{Lise}.
\begin{figure}[htb]
	\includegraphics[height=3cm]{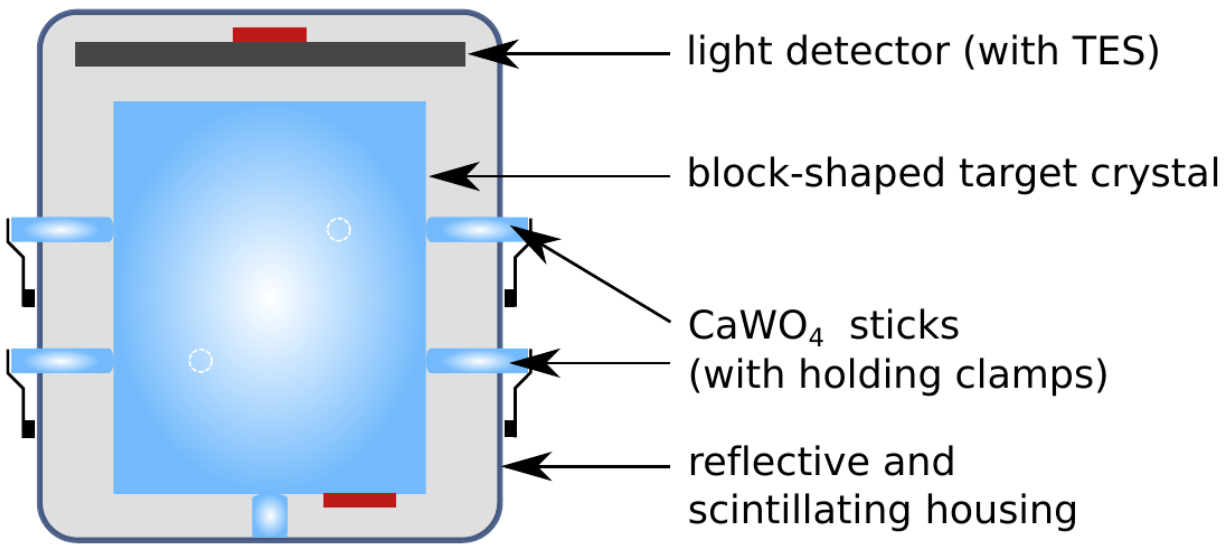}
	\hspace{0.5cm}
	\includegraphics[height=3cm]{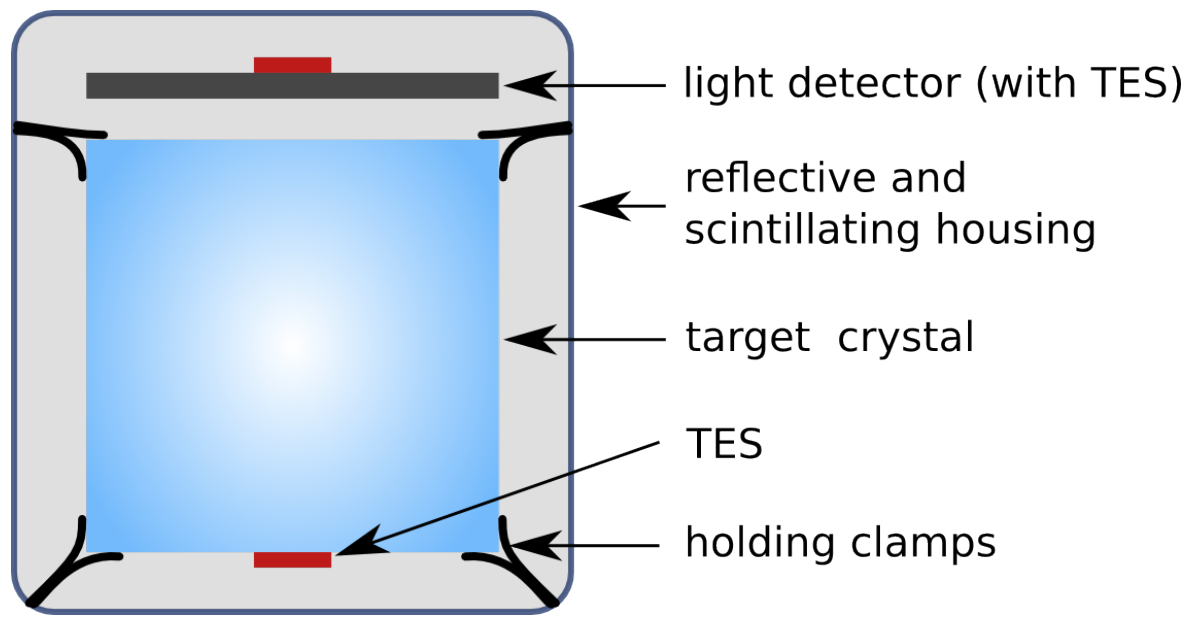}
	\caption{Schematic drawings of the setup of the detector modules TUM40 (left) and Lise (right).}
	\label{fig:Schematics}
\end{figure}
Figure~\ref{fig:Schematics} shows schematic drawings of both detector modules.

For TUM40 we publish a data set which has been used for a dark-matter search \cite{TUM40} as well as background studies \cite{CresstDetectors_1},\cite{TUM40_BG}. The published data set for Lise was used for several dark-matter searches: elastic spin-independent scattering \cite{Lise}, momentum-dependent scattering \cite{Momentum_DM}, and dark-photon dark-matter \cite{DarkPhoton}. Both published data sets contain only events which survive all cuts.
\begin{table}[htb]
	\centering
	\begin{tabular}{|l|l|l|}
		\hline
		& \textbf{TUM40} & \textbf{Lise} \\
		\hline
		Exposure [kg-days] & 29.35 & 52.15 \\
		\hline
	\end{tabular}
	\caption{Exposures before cuts for the published data sets.}
	\label{tab:Exposures}
\end{table}
The exposures before cuts are listed in table~\ref{tab:Exposures} for both data sets.

The light-yield $LY$, i.e., the ratio of the measured light and phonon signals can be used to distinguish electron and nuclear-recoil events, where electron-recoil events originate in interactions of the incident particle with the electrons of CaWO$_4$ and nuclear-recoil events in interactions with the nucleons of CaWO$_4$, respectively.

Since the light yield is related to the generation and detection of photons, the light-yield distribution should be given by a Poisson distribution\footnote{This Poisson distribution has to be convolved with a Normal distribution to account for the baseline noise of the detectors.}. However, if the number of generated and observed photons is large, the Poisson distribution is very well approximated by a Normal distribution. For small numbers of generated and observed photons, where this approximation would fail, the light-yield distribution is dominated by the normally distributed baseline noise of phonon and light detectors. Thus, the assumption of normally distributed light yields can be used for the full energy range \cite{TUM40},\cite{Lise}.

\begin{figure}[htb]
	\includegraphics[width=0.48\textwidth]{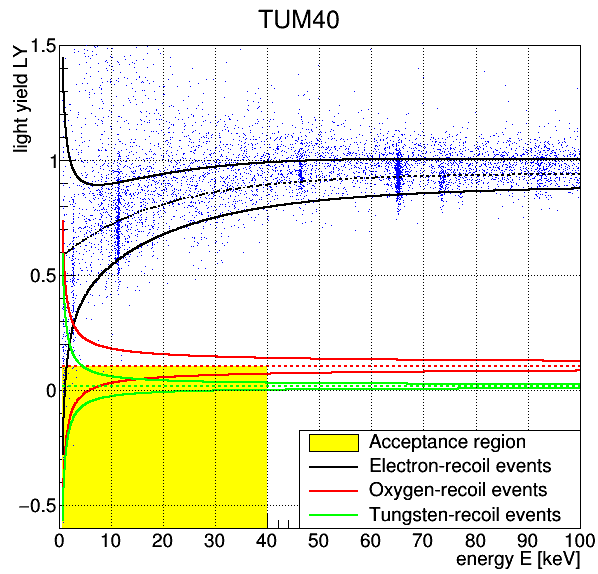}
	\includegraphics[width=0.48\textwidth]{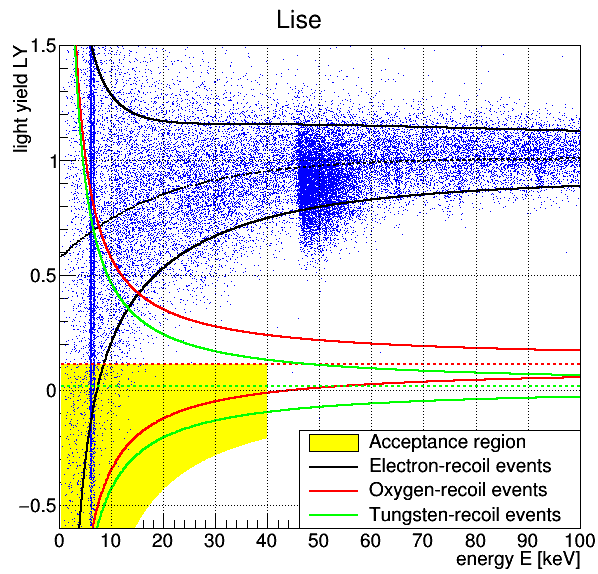}
	\caption{Electron (black), oxygen (red), and tungsten-recoil bands (green) for TUM40 (left) and Lise (right). The dotted lines are the means of the bands; the solid lines depict the central 80\,\% bands. The $\alpha$ and calcium-recoil bands are not shown for reasons of clarity.}
	\label{fig:Bands}
\end{figure}
Figure~\ref{fig:Bands} shows the central 80\,\% bands, i.e., the regions where 80\,\% of the events are expected, for electron, oxygen, and tungsten-recoil events. Above the electron-recoil bands events are visible which are not compatible with the Gaussian shape of the electron-recoil band. These events are called excess-light events \cite{TUM40_BG},\cite{DarkPhoton}. Their origin is not fully understood. However, a likely explanation are external $\beta$s producing additional scintillation light when traversing the scintillating (and reflective) detector housing.

\section{Comparison of data and models}

For dark-matter searches and other studies the distribution of measured events is compared to different models $p_{\text{model}}(E)$ to obtain or constrain parameters of these models. However, a model $p_{\text{model}}(E)$ has to be corrected for finite energy resolution, energy threshold, and cut-survival probability to be comparable to the measured energy distribution:
\begin{equation}\label{equ:Distribution}
	p(E) =   \Theta(E - E_{\text{thr}})\cdot \epsilon_x(E) \cdot \int_0^{\infty} p_{\text{model}}(E') \cdot \mathcal{N}(E-E', \sigma_p^2) dE'
\end{equation}
where the Heaviside step-function $\Theta(E - E_{\text{thr}})$ accounts for the energy threshold and $\epsilon_x(E)$ is the cut-survival probability for event type $x$. To account for finite energy resolution of the phonon detector the model $p_{\text{model}}(E)$ has to be convolved with a normal distribution $\mathcal{N}$ with width $\sigma_p$ being the resolution of the phonon detector.

Equation (\ref{equ:Distribution}) is a simplification which is perfectly valid for energies well above of the energy threshold. For energies not exceeding a distance of one to two times the baseline resolution below the threshold ($\sim 0.45$\,keV for TUM40 and $\sim 0.2$\,keV for Lise) equation (\ref{equ:Distribution}) is still a very good approximation. The simplification of the correct handling of detector resolutions and survival probabilities allows studies of a variety of models with different energy distributions while introducing only a small inaccuracy for very low recoil energies. 

\begin{table}[htb]
	\centering
	\begin{tabular}{|l|l|l|}
		\hline
		\textbf{Parameter} & \textbf{TUM40} & \textbf{Lise} \\
		\hline
		\hline
		$E_{\text{thr}}$\,[keV] & $0.603 \pm 0.002$ & $0.307 \pm 0.004 $\\
		\hline
		$\sigma_p$\,[keV] & $0.090 \pm 0.010$ & $0.062 \pm 0.001$ \\
		\hline
	\end{tabular}
	\caption{Energy threshold $E_{\text{thr}}$ and resolution $\sigma_p$.}
	\label{tab:threshold}
\end{table}
Table~\ref{tab:threshold} lists the energy resolutions and thresholds for TUM40 \cite{TUM40} and Lise \cite{Lise}. In general, the energy resolution depends on the energy. The values for $\sigma_p$ given in table~\ref{tab:threshold} are the baseline resolutions, i.e., the resolutions for zero energy deposition. The assumption of constant energy resolution is valid for low energies of $\mathcal{O}$(keV). If required, the energy dependency of the resolution for larger energies could be obtained from several lines within the electron-recoil band.

The cut-survival probability $\epsilon_x(E)$ can be expressed as:
\begin{equation}
	\epsilon_x(E) = f_x(E) \cdot \epsilon_0(E)
\end{equation}
where $\epsilon_0(E)$ is the survival probability for all cuts except a selection on the light yield and $f_x(E)$ is the fraction of events of type $x$ accepted by the light-yield selection. The survival probability $\epsilon_0(E)$ is determined by applying all cuts to artificial pulses with several (closely spaced) discrete energies. See \cite{TUM40},\cite{Lise} for further details. The full cut-survival probabilities are also published for electron and nuclear-recoil events within the electron-recoil band and the region of interest for dark-matter search, respectively.

\section{Published data-files}

The published data from CRESST-II Phase 2 are stored in several ASCII files. The first lines starting with ``\#'' contain a few comments about the contents of each file. The following lines contain the energy of one event per line.

\subsection{Electron-recoil band}

The files \verb=TUM40_eRecoils.dat= and \verb=Lise_eRecoils.dat= contain only events within the electron-recoil band depicted in black in figure~\ref{fig:Bands}.
\begin{figure}[htb]
	\includegraphics[width=0.48\textwidth]{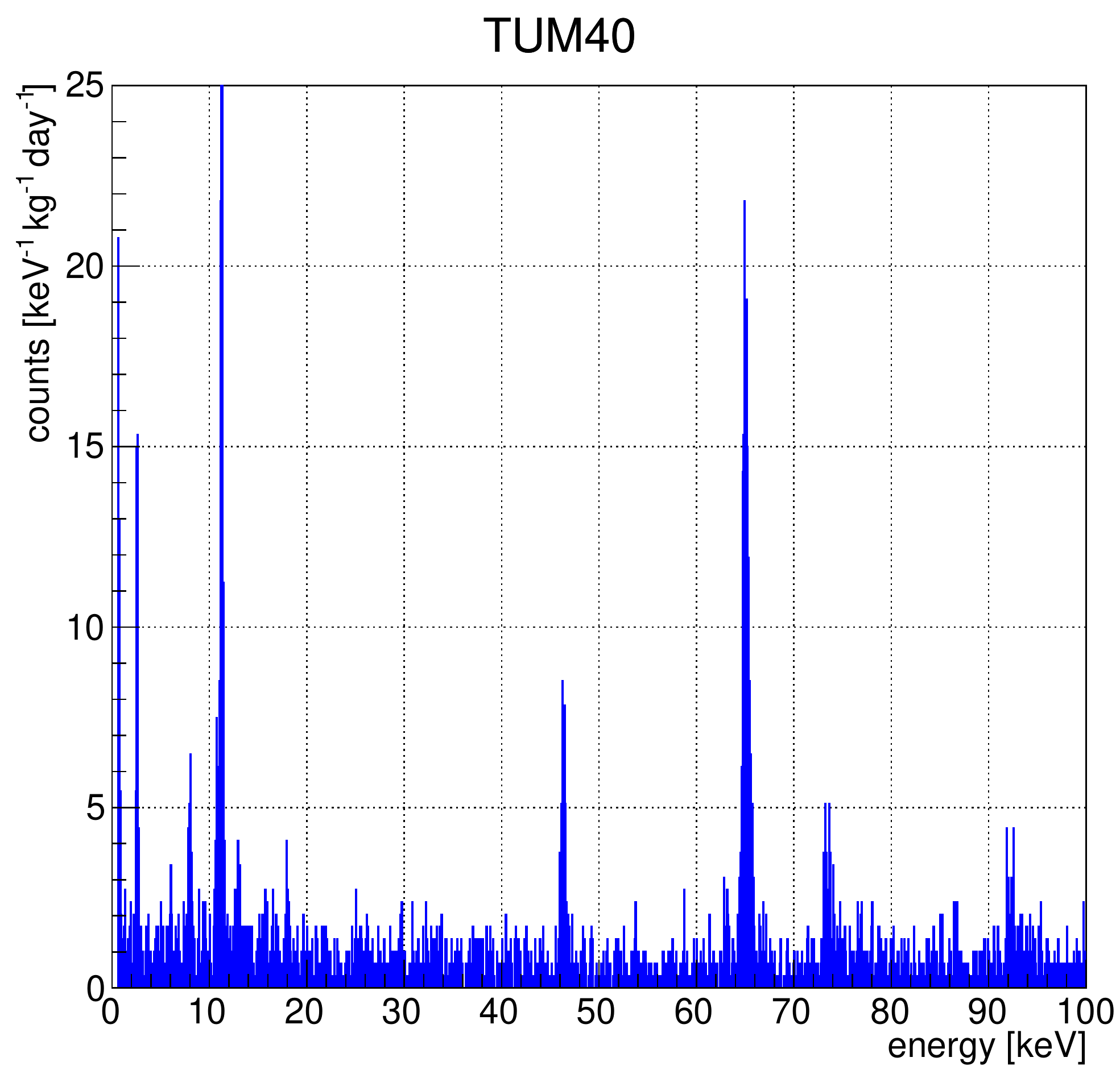}
	\includegraphics[width=0.48\textwidth]{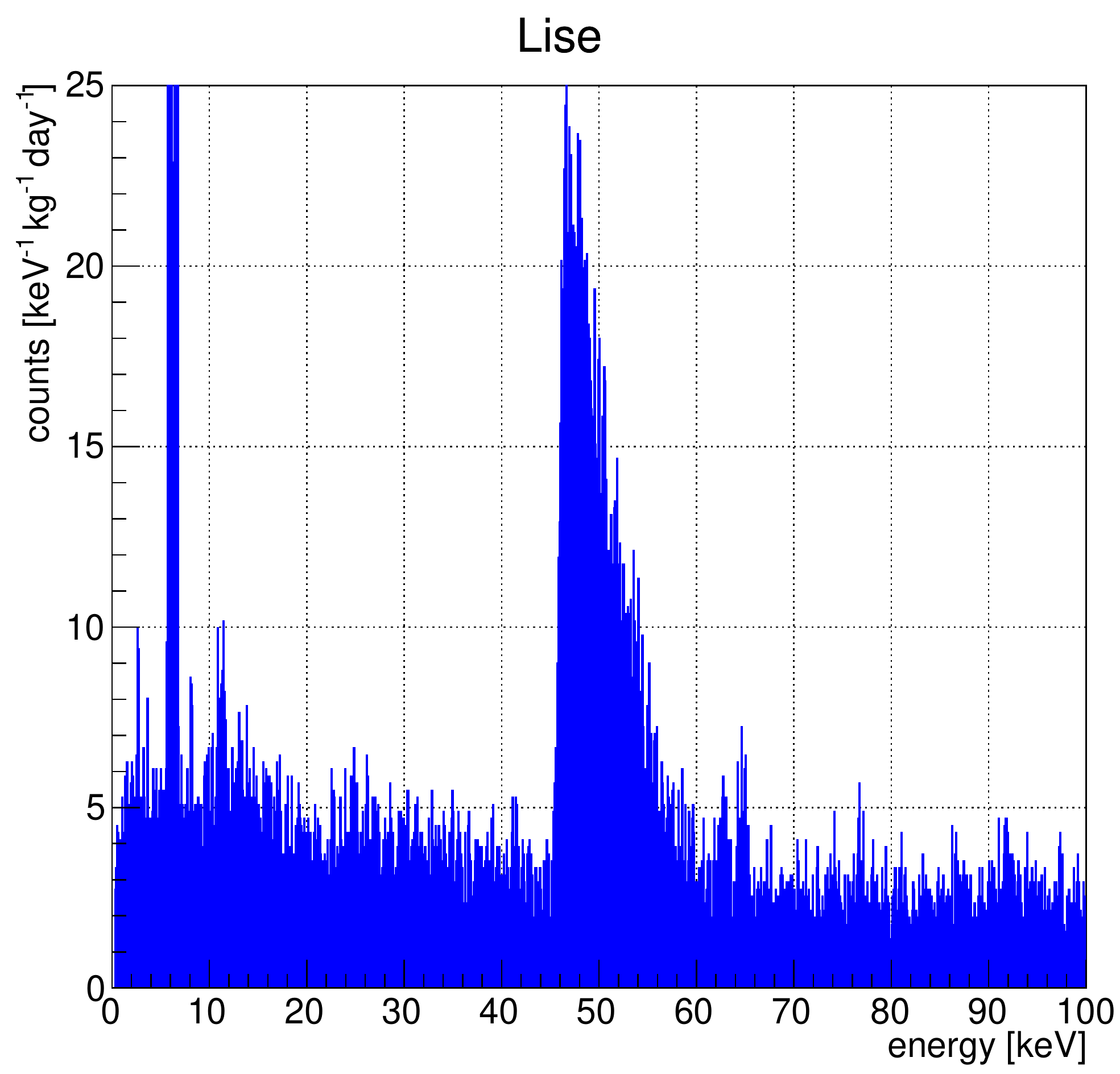}
	\caption{Energy spectra for electron-recoil events for TUM40 (left) and Lise(right).}
	\label{fig:ElectronRecoil}
\end{figure}
Figure~\ref{fig:ElectronRecoil} shows the energy spectra of the electron-recoil events for TUM40 (left) and Lise (right). See \cite{CresstDetectors_1}-\cite{TUM40_BG}, respectively, for a detailed description of the features of the spectra.

\begin{figure}
	\includegraphics[width=0.48\textwidth]{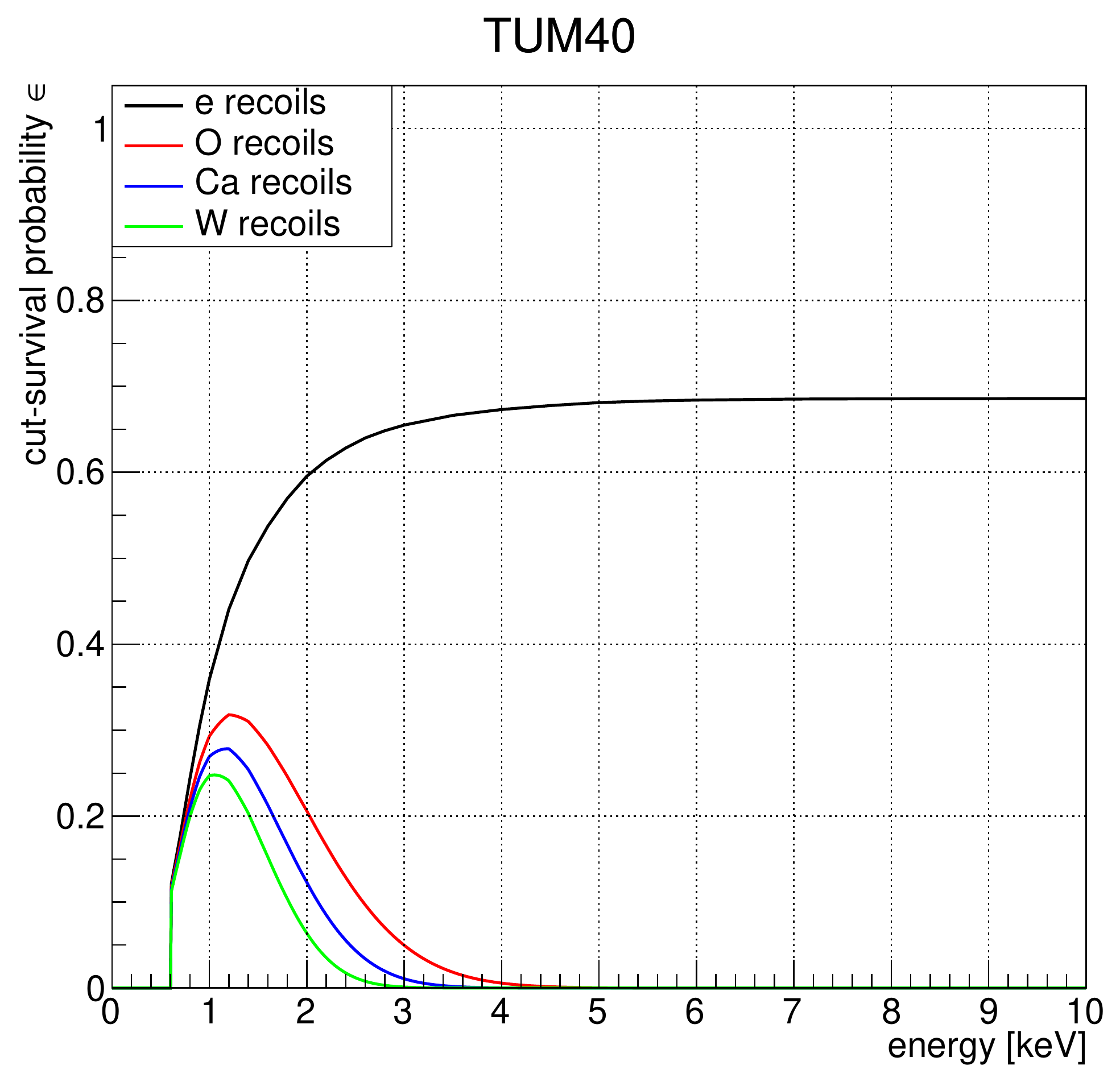}
	\includegraphics[width=0.48\textwidth]{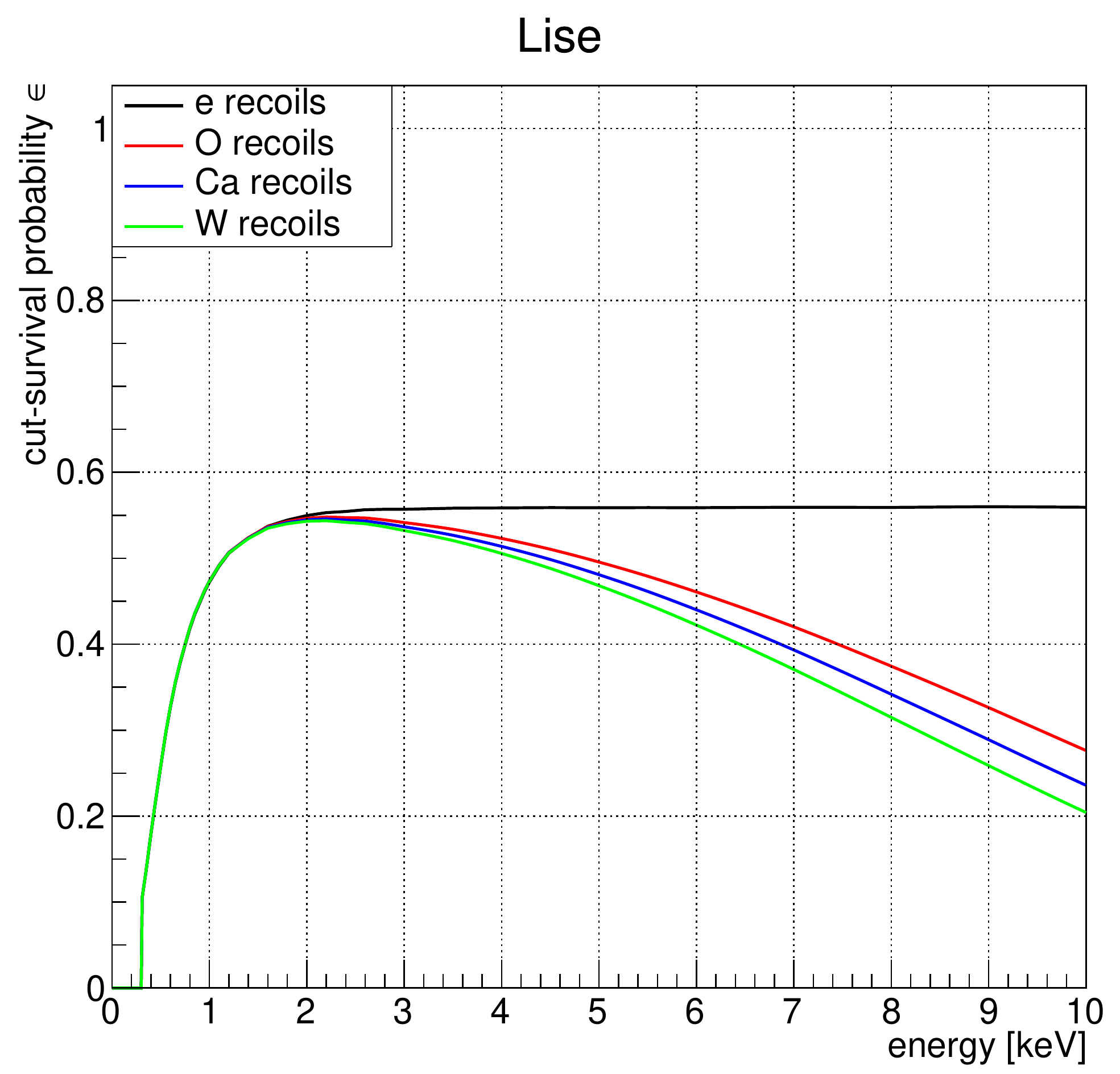}
	\caption{Cut-survival probabilities for TUM40 (left) and Lise (right) for electron and nuclear-recoil events within the electron-recoil band.}
	\label{fig:eff_gamma}
\end{figure}
The cut-survival probabilities for events within the electron-recoil band shown in figure~\ref{fig:eff_gamma} are also published as ASCII files:
\begin{itemize}
  \item \verb=TUM40_eff_eRecoils_e.dat= and \verb=Lise_eff_eRecoils_e.dat= for electron-recoil events (black line in figure~\ref{fig:eff_gamma})
  \item \verb=TUM40_eff_eRecoils_O.dat= and \verb=Lise_eff_eRecoils_O.dat= for oxygen-recoil events (red line in figure~\ref{fig:eff_gamma})
  \item \verb=TUM40_eff_eRecoils_Ca.dat= and \verb=Lise_eff_eRecoils_Ca.dat= for calcium-recoil events (blue line in figure~\ref{fig:eff_gamma})
  \item \verb=TUM40_eff_eRecoils_W.dat= and \verb=Lise_eff_eRecoils_W.dat= for tungsten-recoil events (green line in figure~\ref{fig:eff_gamma})
\end{itemize}
Similar to the event-data files, the first lines starting with ``\#'' contain a few comments about the contents of each file. The following lines contain the cut survival probabilities arranged in two columns:
\begin{enumerate}
  \item The energy $E$ in keV.
  \item The full cut-survival probability $\epsilon_x(E)$\footnote{An survival probability of one corresponds to the exposures given in table~\ref{tab:Exposures}} for the energy in column 1. 
\end{enumerate}
The cut-survival probabilities have to be extrapolated for energies not listed in the files. For energies larger than the listed energies, the cut-survival probabilities remain constant.

The cut-survival probabilities for Lise have a small dip ($\lesssim 1$\,\%) around 30\,keV which is related to all cuts being optimized for small energies below 10\,keV. Due to its smallness, the dip does not impact any studies of Lise data.

\subsection{Acceptance region for dark-matter search}

The files \verb=TUM40_AR.dat= and \verb=Lise_AR.dat= contain only events within the acceptance regions for dark-matter search as defined in \cite{TUM40} and \cite{Lise}, respectively. The acceptance regions for TUM40 and Lise are slightly different. For both data sets only events with energies inside the interval $[E_{\text{thr}}, 40\text{\,keV}]$ are accepted. The upper boundary for the light yields is also similar for both data sets: The mean of the oxygen band. The lower boundary for the light yield, however, is different for TUM40 and Lise. For Lise the lower boundary is the 0.5\,\% quantile of the tungsten band \cite{Lise}. For TUM40 there is no lower boundary, i.e., all events with light yields smaller than the mean of the oxygen band are accepted for dark-matter searches \cite{TUM40}. It is important to mention that this difference in the definitions of the acceptance regions has no impact for the dark-matter sensitivity of CRESST-II.

\begin{figure}[htb]
	\includegraphics[width=0.48\textwidth]{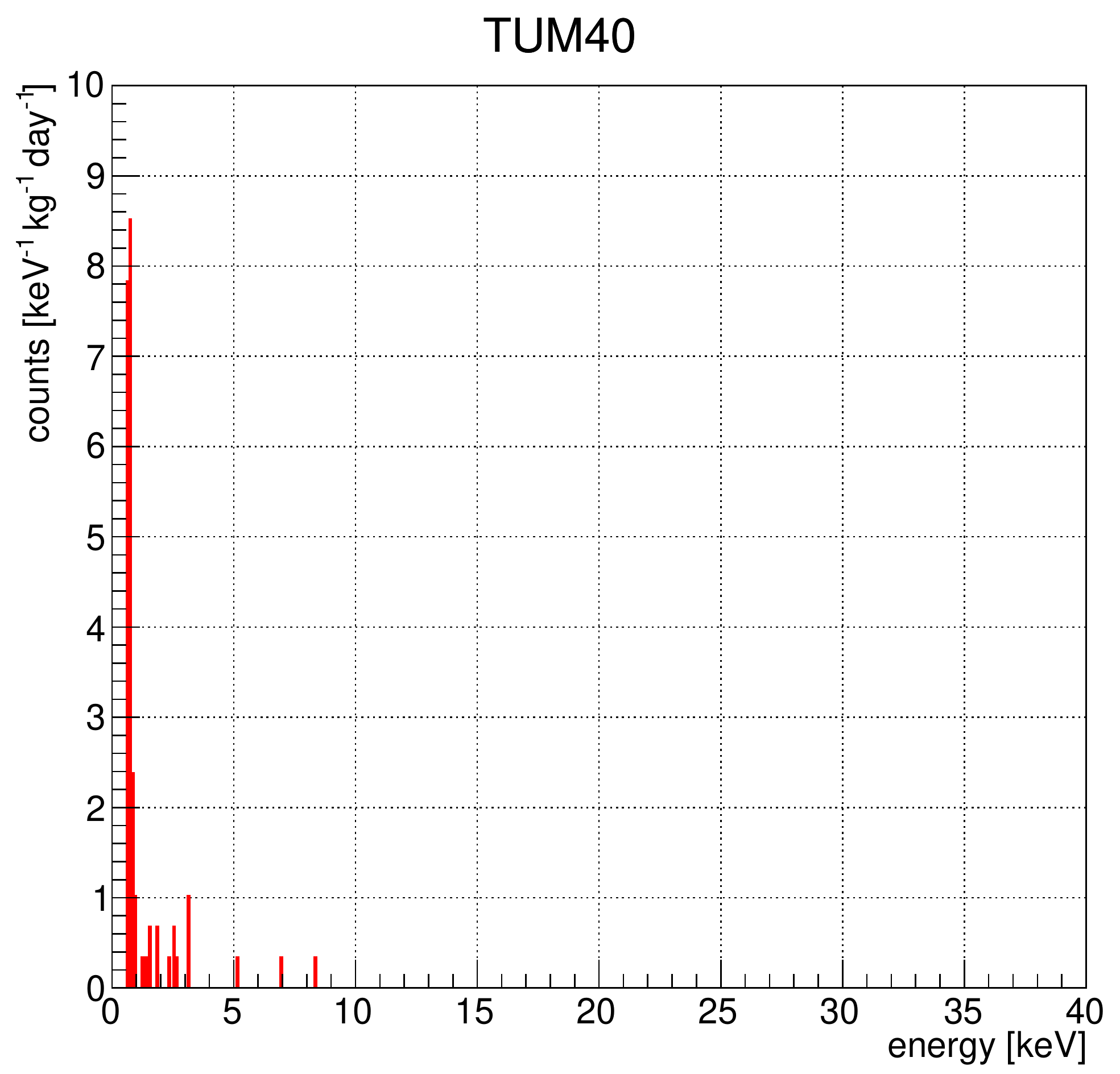}
	\includegraphics[width=0.48\textwidth]{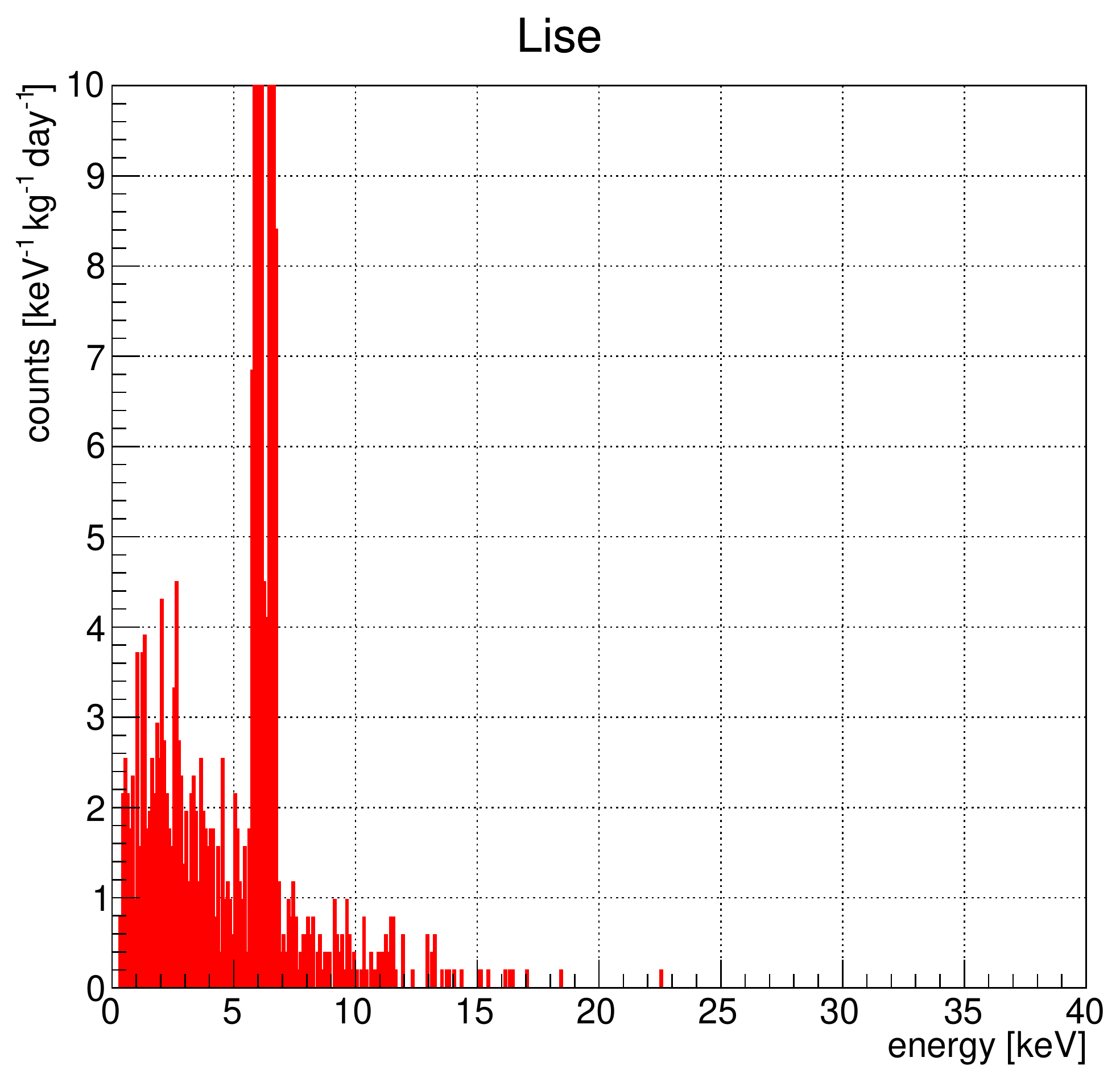}
	\caption{Energy spectra of the events with in the acceptance regions for TUM40 (left) and Lise (right).}
	\label{fig:ROI}
\end{figure}
Figure~\ref{fig:ROI} depicts the energy spectra of the events within the acceptance regions. The majority of these events is compatible with leakage from the electron-recoil band to the acceptance region. Further details on these spectra and how they are used for dark-matter searches can be found in \cite{TUM40},\cite{Lise},\cite{Momentum_DM}.

\begin{figure}
	\includegraphics[width=0.48\textwidth]{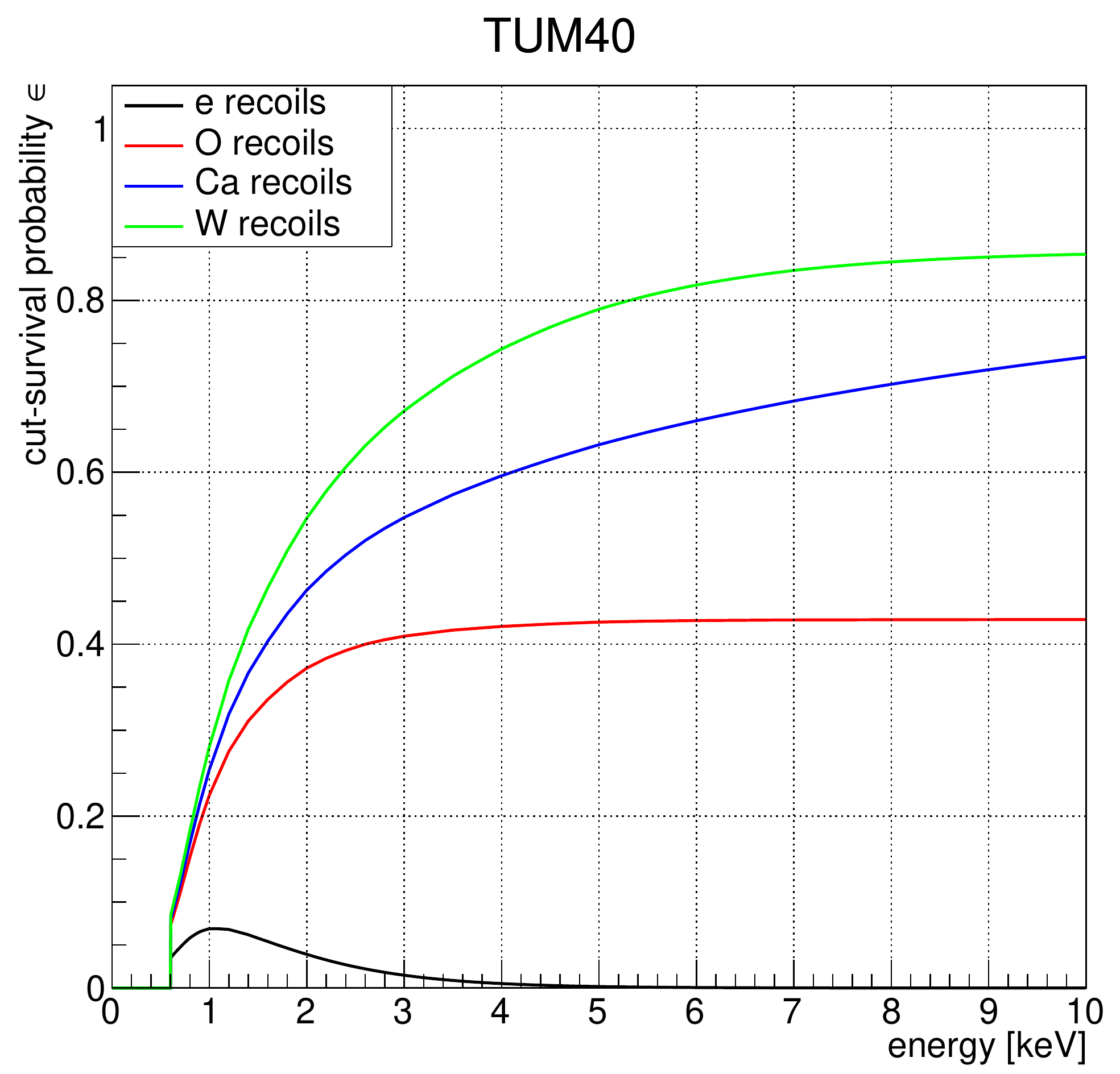}
	\includegraphics[width=0.48\textwidth]{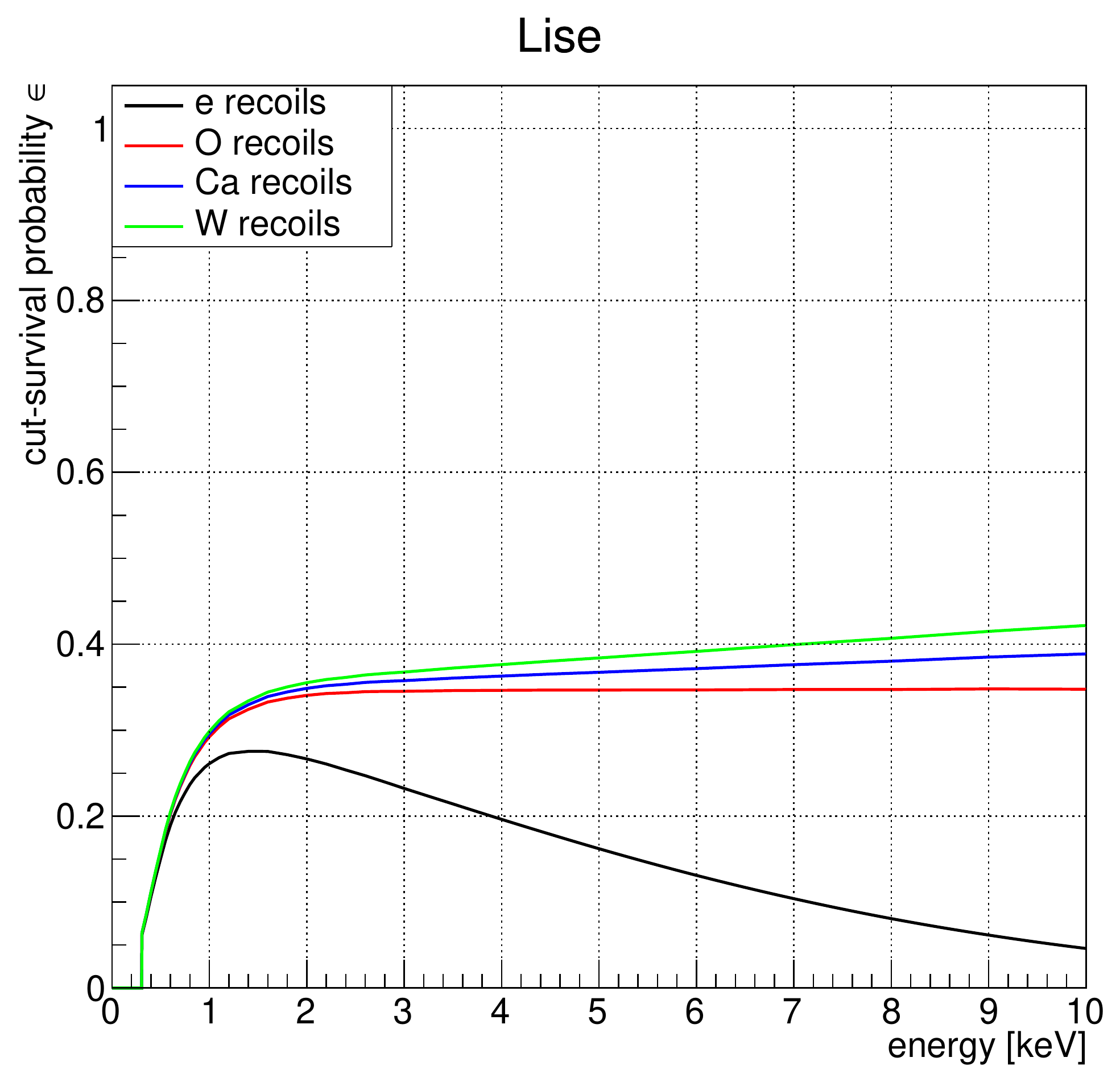}
	\caption{Cut-survival probabilities for TUM40 (left) and Lise (right) for electron and nuclear-recoil events within the acceptance regions.}
	\label{fig:eff_ROI}
\end{figure}
The cut-survival probabilities for events within the acceptance regions shown in figure~\ref{fig:eff_gamma} are also published as ASCII files:
\begin{itemize}
  \item \verb=TUM40_eff_AR_e.dat= and \verb=Lise_eff_AR_e.dat= for electron-recoil events (black line in figure~\ref{fig:eff_ROI})
  \item \verb=TUM40_eff_AR_O.dat= and \verb=Lise_eff_AR_O.dat= for oxygen-recoil events (red line in figure~\ref{fig:eff_ROI})
  \item \verb=TUM40_eff_AR_Ca.dat= and \verb=Lise_eff_AR_Ca.dat= for calcium-recoil events (blue line in figure~\ref{fig:eff_ROI})
  \item \verb=TUM40_eff_AR_W.dat= and \verb=Lise_eff_AR_W.dat= for tungsten-recoil events (green line in figure~\ref{fig:eff_ROI})
\end{itemize}
The format for these files is identical to the files containing the cut-survival probabilities for the electron-recoil band.

Again, there is a small dip around 30\,keV in the cut-survival probabilities of Lise.

\section{Test case}

\begin{figure}
	\centering
	\includegraphics[width=0.8\textwidth]{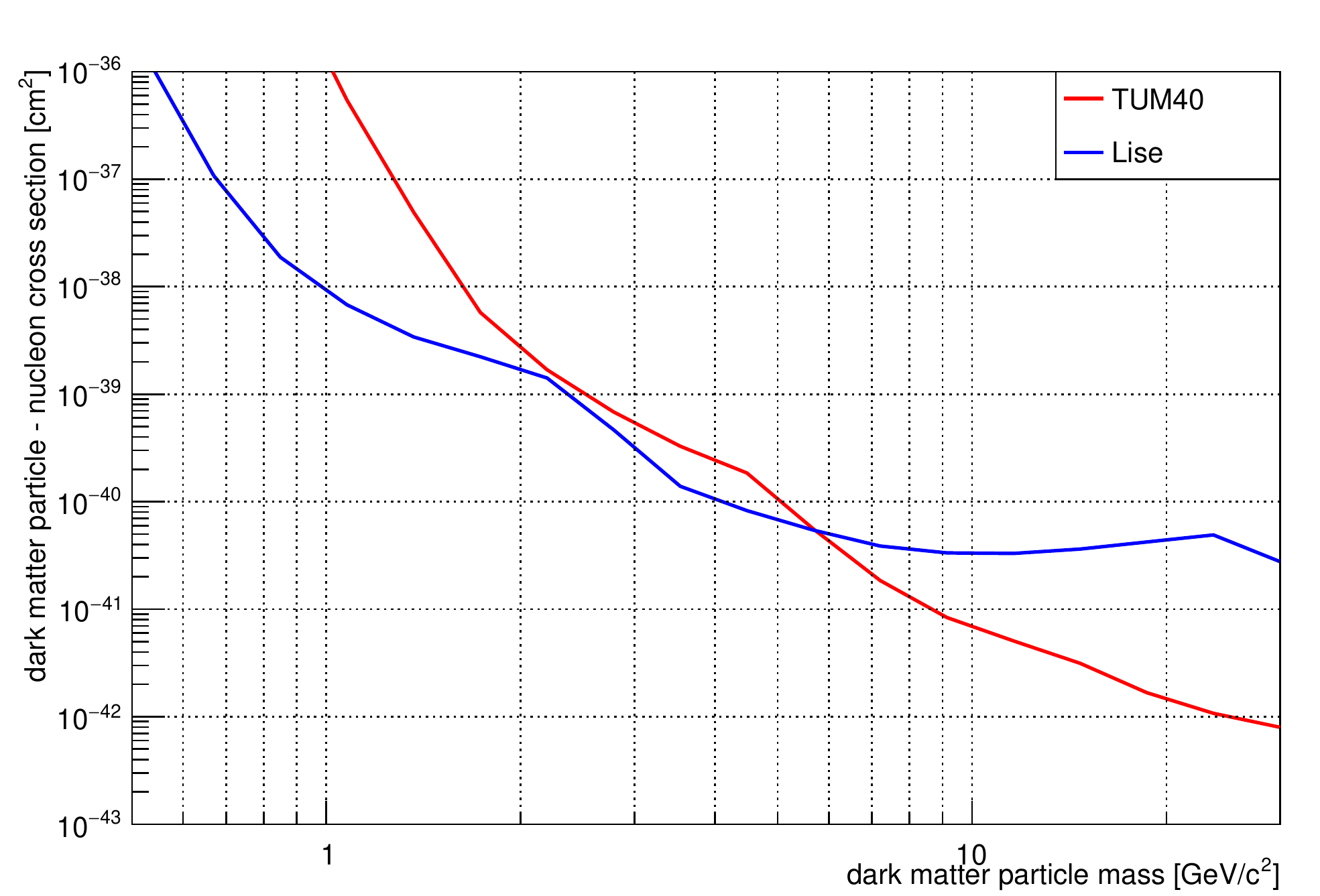}
	\caption{Test cases: Upper 90\,\% exclusion limits for the cross section of spin-independent elastic scattering of dark-matter particles of nuclei for TUM40 and Lise.}
	\label{fig:TestCases}
\end{figure}
The exclusion limits depicted in figure~\ref{fig:TestCases} were calculated by applying Yellin's optimum interval method \cite{Yellin} to the data sets supplied with this document. The files \verb=TUM40_TestCase.dat= and \verb=Lise_TestCase.dat= contain the data points for both limits. As for the other files the first lines starting with ``\#'' contain some information about the contents of the files. The following lines contain the data points in two columns: The first column contains the mass of dark-matter particles in GeV/c$^2$ and the second column contains the exclusion limit in cm$^2$.

Following the energy constraints of equation (\ref{equ:Distribution}), we calculated the exclusion limits only for energies above 0.8\,GeV/c$^2$ for TUM40 and 0.5\,GeV/c$^2$ for Lise, respectively. The dark-matter particles with the minimum masses generate recoil energies up to $0.45$ and $0.2$\,keV, respectively. Thus, for smaller masses the full energy distributions would not be covered by equation (\ref{equ:Distribution}).

\section{Citation}

If you base your work on our data, we kindly ask to cite this document as well as \cite{TUM40} for TUM40 data and \cite{Lise} for Lise data, respectively.

\end{document}